# Giant-Shell CdSe/CdS Nanocrystals: Exciton Coupling to Shell Phonons Investigated by Resonant Raman Spectroscopy

*Miao-Ling Lin[1,2], Mario Miscuglio[3], Anatolii Polovitsyn[4], Yuchen Leng[1,2], Beatriz Martín-García[3], Iwan Moreels[4], Ping-Heng Tan[1,2] and Roman Krahne[3]\**,

[1] State Key Laboratory of Superlattices and Microstructures, Institute of Semiconductors, Chinese Academy of Sciences, Beijing, 100083, China

[2] College of Materials Science and Opto-Electronic Technology & CAS Center of Excellence in Topological Quantum Computation, University of Chinese Academy of Sciences, Beijing 100190, China

[3] Istituto Italiano di Tecnologia (IIT), Via Morego 30, 16163 Genoa, Italy

[4] Department of Chemistry, Ghent University, Krijgslaan 281-S3, 9000 Gent, Belgium

Corresponding author Email: Roman.krahne@iit.it

ABSTRACT

The interaction between excitons and phonons in semiconductor nanocrystals plays a crucial role in the exciton energy spectrum and dynamics, and thus in their optical properties. We investigate the exciton-phonon coupling in giant-shell CdSe/CdS core-shell nanocrystals via resonant Raman spectroscopy. The Huang-Rhys parameter is evaluated by the intensity ratio of the longitudinal-optical (LO) phonon of CdS with its first multiscattering (2LO) replica. We used four different excitation wavelengths in the range from the onset of the CdS shell absorption to well above the CdS shell band edge to get insight into resonance effects of the CdS LO phonon with high energy excitonic transitions. The isotropic spherical giant-shell nanocrystals show consistently stronger exciton-phonon coupling as compared to the anisotropic rod-shaped dot-in-rod (DiR) architecture, and the 2LO/LO intensity ratio decreases for excitation wavelengths approaching the CdS band edge. The strong exciton-phonon coupling in the spherical giant-shell nanocrystals can be related to the delocalization of the electronic wave functions. Furthermore, we observe the radial breathing modes of the GS nanocrystals and their overtones by ultra-low frequency Raman spectroscopy with nonresonant excitation, using laser energies well below the band gap of the heteronanocrystals, and highlight the differences between higher order optical and acoustic phonon modes.

Colloidal nanocrystals present an interesting material for light emission since their optical properties can be carefully tuned by size, shape and composition.[1-3] In particular, CdSe nanocrystals constitute a model system that has been extensively studied, as substantial shape control has been achieved together with a thorough understanding of their exciton level structure.[4-6] CdSe/CdS core-shell nanocrystals manifest improved emission properties due to the epitaxial cladding of the core with the higher band gap material CdS. Spherical dot-in-dot[7-8] and anisotropic dot-in-rod[9-10] and rod-in-rod[11-12] core-shell architectures have been exploited for applications in light emission, where high photoluminescence quantum yield and control over the photophysical properties are desired. The specific band structure, with a small offset in the conduction band and a large offset in the valence band leads to a quasi-type II heterostructure with different localization volume for electrons and holes.[13] Recently, core-shell CdSe/CdS heterostructures with a significantly increased thickness of the CdS shell have been realized, which are typically referred to as "giant shell" (GS)[14-16] or thick-shell nanocrystals.[17-18] The much thicker CdS shell leads to a stronger delocalization of the electron wave function into the shell material, while the hole wave function remains localized to the CdSe core.[15, 19-20]

The exciton-phonon coupling plays an important role in the relaxation processes of high-energy excitons, in the emission of the band edge excitons,[21] and in the dissipation of thermal energy in colloidal nanocrystal materials. For CdSe nanocrystals the coupling to longitudinal-optical (LO) phonons has been studied experimentally (by Raman spectroscopy[22] and fluorescence line narrowing (FLN)[23]) and theoretically in great detail.[24-26] In an adiabatic approximation, the coupling of the electronic charge with the fields induced by the vibrational motion of a polar crystal, via the Fröhlich interaction, can be considered by a displaced harmonic oscillator model within the Franck-Condon model,[24, 27] which allows to relate the intensities of different scattering orders of the LO phonon mode to the displacement factor in the above mentioned harmonic oscillator model. Band mixing can lead to non-adiabatic phonon assisted transitions, which were considered in CdSe and PbS nanocrystals in ref. [28]. Inelastic neutron scattering studies have shown that the mechanical softening at the surfaces of nanocrystals favors the excitation of multiphonon processes.[29] Experimentally, the exciton-phonon coupling strength can be assessed by Raman spectroscopy, evaluating the Huang-Rhys parameter obtained by the ratio of the intensities of the different scattering orders of the LO phonon resonance.[24] For a stringent relation of the intensity ratio of the LO phonon scattering orders to the exciton-phonon coupling, also the broadening of the involved electronic transitions has to be considered.[30] In core-shell nanocrystals, the behavior becomes more complicated due to possible coupling to LO phonons from the core and from the shell materials, due to different electron and hole wave function distributions, and due to a non-trivial dependence on the excitation wavelength.[31-34] In general, one could expect that the reduced overlap between electron and hole wave functions, and concomitant enhanced polarizability, should lead to an enhancement of the exciton-phonon coupling, which indeed was observed for type II CdTe/CdSe heteronanocrystals.[35] In type I and quasi-type II CdSe/ZnS and CdSe/CdS core-shell nanocrystals the LO phonon excitations and the exciton-phonon coupling have been studied by Raman spectroscopy and FLN.[36-39] Excitation above the CdS bandgap led mainly to coupling with the CdS phonons, since the high-energy exciton states involved in the Raman scattering process should be delocalized over a large part of the CdS volume.[40]

In this work, we investigate exciton-phonon coupling in wurtzite giant-shell nanocrystals [41] by resonant Raman spectroscopy at energies around and above the CdS band gap, and compare the results with dot-in-rod heteronanocrystals synthesized following a standard literature protocol[10]. We evaluate the Huang-Rhys parameter $S$ by the intensity ratio of the CdS LO phonon with its first multiphonon replica using resonant Raman spectroscopy, exciting the samples at wavelengths from 442 nm to 502 nm, which corresponds to energies above and at the CdS absorption edge of the nanocrystals. We observe strong higher order replicas of the CdS LO phonon that can be well-fitted with Lorentzian peaks. The 2LO/LO intensity ratio, I(2LO)/I(LO), obtained from this fitting, is significantly larger in GS nanocrystals as compared to DiRs. Interestingly, we find that for comparatively thin shells in the core/shell architecture, of ~1.5 nm thickness, the intensity ratio is significantly reduced. For thicker shells, the intensity ratio remains more or less constant with respect to geometrical parameters such as nanocrystal diameter, volume, or core size. For all core-shell samples the I(2LO)/I(LO) ratio decreases when the excitation wavelength approaches the CdS band gap energy, which is the opposite trend compared what we observe from monomaterial CdS structures. In principle, the strong exciton-phonon coupling in GS nanocrystals should also be reflected in the spectra of the radial breathing modes (RBMs). To avoid the strong photoluminescence background present under resonant conditions, the acoustic phonon modes were obtained with nonresonant excitation at 785 nm, well below the nanocrystal band edge. The fundamental RBM modes and their higher harmonics can be well resolved, and their frequencies can be well approximated by applying Lamb theory for oscillations in a continuous medium with the sound velocity as a fitting parameter. Under such nonresonant excitation, the Raman scattering cross sections fall by orders of magnitude, and we no longer discerned clear 2LO peaks, hence analysis of the LO peak intensity ratios could not be used to evaluate the exciton-phonon coupling. Furthermore, the higher-order RBM resonances in the Raman spectra are likely to be true higher harmonics of the fundamental vibrational oscillation in the nanocrystals, and therefore cannot be associated to multiphonon scattering events, as it is the case for the LO phonon replicas. This intrinsically hampers the analysis of the coupling strength of excitons to acoustic phonons via higher order peak intensities.

Figure 1a shows representative absorption and emission spectra of the GS and DiR samples. The absorption due to the CdS shell manifests a strong onset around 490 – 500 nm, with a steeper slope and onset at higher energy for the DiRs. However, for GS nanocrystal samples with different core size and different overall diameter, the CdS absorption band edge is very similar since for such sizes confinement effects are absent. The emission of the core-shell heterostructures is dominated by transitions related to the CdSe cores, and is more blue-shifted for the DiRs due to increased quantum confinement. The excitation wavelengths for the resonant Raman measurements were chosen at and above the CdS band edge of the nanocrystals, as indicated by the dashed vertical lines that mark wavelengths at 502, 488, 457, and 442 nm. To detect the RBMs, nonresonant Raman spectra were measured at 785 nm (red dashed line). The reason for nonresonant excitation of the acoustic modes is that otherwise the nanocrystal emission camouflages the vibrational signal. The energy band scheme of the core-shell architecture is illustrated in the inset: a quasi-type II band alignment occurs for excitons at the band edge due to the stronger confinement of the holes to the core region (see red dotted lines). We note that for equal penetration length of the electron wave function into the CdS volume, the delocalization volume is much larger in the GS nanocrystals due to their spherical symmetry (as opposed to DiRs that maintain strong electron confinement in the radial direction of the rod-shape). Figure 1b shows the Raman spectra of GS nanocrystals (blue circles) with 14 nm diameter and 5 nm core size in the range from 150-1000 cm$^{-1}$, where the LO band of CdS is evident, together with strong 2LO and 3LO replicas. To analyze the spectra, the data was fitted with Lorentzian peaks, as detailed in the SI. Both frequency and peak width scale linearly with the scattering order of the LO phonon. The narrow peak width and the linear scaling of the higher scattering orders in frequency point to scattering of LO phonons at the center of the Brillouin zone, *i.e.* with wave vector $k$ near zero, because contributions with higher $k$ should lead to a broadened LO phonon peak and to slightly redshifted higher scattering orders. The linear scaling of the peak width of the higher scattering orders agrees very well with the population relaxation mechanism discussed in ref. [42], which indicates that the higher scattering orders can be related to the same excitonic transition. Therefore, a relation of I(2LO)/I(LO) to the Huang Rhys factor should be valid.[30] We note all GS nanocrystal samples in this study follow this linear scaling high good accuracy for excitations at 457, 488, and 502 nm, as reported in Table S1. At 442 nm, the full-width at half-maximum (FWHM) of the 2LO peak tends to be slightly broader than that the LO one, however, restricting it to twice the FWHM of the LO peak also leads to very accurate fitting (please see SI section 2 for more information).

For a comparison of the exciton-phonon coupling in GS and DiR nanocrystals, similar resonance conditions with the electronic levels should be selected. In Figure 1c, we show the Raman spectra of GS and DiR nanocrystals that were excited at a similar position at the onset of their CdS shell absorption, as highlighted by the black rectangle in Figure 1a, which should correspond to similar resonant scattering conditions with respect to the electronic transitions. For the GS nanocrystals, the excitation is at 502 nm, while for the DiRs it is at 488 nm due to the smaller diameter with respect to the GS nanocrystals. To evaluate the electron-phonon coupling, we analyzed the intensity (*i.e.* area) ratio of 2LO and LO Lorentzian peaks, as detailed in the SI. For such similar resonance conditions with the excitonic transitions, the I(2LO)/I(LO) is about a factor of 1.3 larger for the GS nanocrystals with respect to the DiR nanocrystals, which holds across the different core and shell sizes.

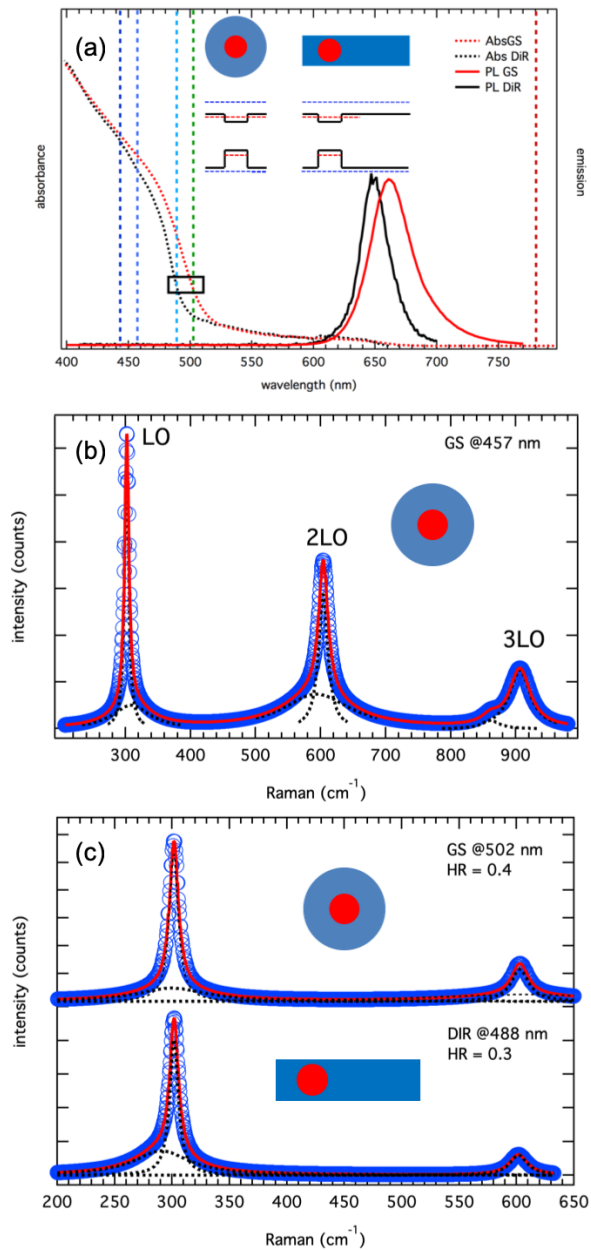

**Figure 1.** (a) Absorption (dashed) and emission (solid) spectra of giant-shell (red, 14 nm diameter, 7 nm core size) and DiR (black, 5.8 nm diameter, 5.7 nm core size) nanocrystal samples recorded in solution. The blue and green vertical dashed lines indicate the excitation wavelength for the resonant Raman spectroscopy measurements (Figure 1b,c and Figure 2), and the red vertical dashed line that of the nonresonant Raman experiments (Figure 3). The inset shows schemes of the core-shell architecture and the correponding band alignment. (b) Raman spectrum of GS nanocrystals with 15 nm overall diameter and 5 nm core diameter recorded at 457 nm excitation wavelength (blue circles), together with Lorentz fitting of the peaks. LO, 2LO and 3LO Raman peaks are observed at 300, 600, and 900 cm$^{-1}$. (c) Raman

spectra recorded at 502 and 488 nm excitation wavelength for GS (14 nm diameter, 7 nm core diameter) and DiR (5.8 nm diameter, 5.7 nm core diameter) nanocrystals, respectively. These laser wavelengths correspond to a similar onset in their absorption spectra, as highlighted by the rectangle in (a).

Typical Raman spectra of DiRs and GS nanocrystals at different excitation wavelengths at and above the CdS band gap are depicted in Figure 2a, together with the Lorentzian fitting of the LO and 2LO bands. We show heterostructures with the same CdSe core size of 4 nm, where the GS nanocrystals had a diameter of 13 nm, and the DiRs 4.8 nm diameter and 30 nm length. For all excitation wavelengths, the 2LO peak is much stronger for the GS nanocrystal samples. Figure 2b shows the normalized resonant Raman spectra (at 442 nm) of GS nanocrystal samples with similar core size around 4 nm, but different shell thickness. Here the GS nanocrystals with 13 nm and 11 nm total diameter show nearly identical behaviour, while the one with 7 nm total diameter manifests a significantly weaker 2 LO peak. Therefore, there is a limit on minimum shell thickness, of ~2-3 nm, for the strong GS phonon coupling behaviour, and above this limit the coupling does not depend critically anymore on the CdS shell volume. Spectra recorded at 457 nm excitation wavelength are reported in Fig. S1 in the SI and show the same trend. The I(2LO)/I(LO) ratios from all experiments are reported in Table S1, and plotted in Figure 2c versus the core volume fraction (*i.e.* the core volume divided by the overall volume of the nanocrystals, $V_{CORE}/V_{TOTAL}$). Excitation well above the CdS shell band gap (at 442 and 457 nm) leads to I(2LO)/I(LO) values in the range from 0.6 – 0.9 for the GS nanocrystals, while for the DiRs we obtained values between 0.3-0.5. Excitation close to the CdS band gap, at 502 nm for the GS and 488 nm for the DiRs, leads to a smaller I(2LO)/I(LO) ratio. From the data in Figure 2c, we see that excitons at high energy couple more efficiently to the CdS LO phonons, which can be understood by the stronger delocalization of these hot excitons in the CdS shell volume. The I(2LO)/I(LO) ratio in CdSe/CdS nanocrystals decreases when the excitation wavelength approaches the CdS shell band gap [32], which can be related to the reduced wavefunction delocalization at lower energies. However, the delocalization of the exciton strongly depends on the geometry. The crucial point is that for a given penetration depth of the electron wave function in the shell material, the volume occupied by the electrons in the spherical giant shell is much larger than in the cylindrical DiR, where axial delocalization is strongly limited by the comparatively small rod diameter. Therefore, the polarizability is much stronger in the GSs with respect to the DiRs, which leads to a stronger exciton-phonon coupling. Thus, the difference in coupling strength is intrinsic to the heteronanocrystal shape and the thick-shell architecture.

The CdSe/CdS core-shell nanocrystals show a completely different exciton-phonon coupling behaviour as has been reported for monomaterial CdSe or CdS nanocrystals. In the monomaterial case, the coupling strength scales with the inverse of the diameter, [22, 36, 43] and increases when the excitation wavelength approaches the band gap[42,44]. In Figures S2-S4 we report the optical and resonant Raman spectra of giant CdS nanocrystals and CdS nanorods that confirm this behavior. Here the giant CdS nanocrystals manifest an I(2LO)/I(LO) ratio of 0.61 at 442 nm, which increases to 0.65 at 457 nm. Excitation on the low-energy shoulder of the photoluminescence peak, at 532 nm, yields a I(2LO)/I(LO) ratio of 0.75. However, this high value could be related to an outgoing resonance that increases the 2LO peak strength, as reported in ref. [45]. Another peculiar difference in the exciton-phonon coupling between CdSe and CdS nanocrystals and their GS core-shell counterparts is that the influence of the soft surface[29] should be negligible in the GS nanocrystals, because the exciton wave functions do not extend to the shell surface, especially for excitation at the heteronanocrystal band edge. We note that the signal from CdSe LO modes is not observed in our spectra with excitation above the CdS band gap [32,40], which is partially due to the resonance conditions, and because the volume of the CdSe cores is much smaller than that of the CdS shell.

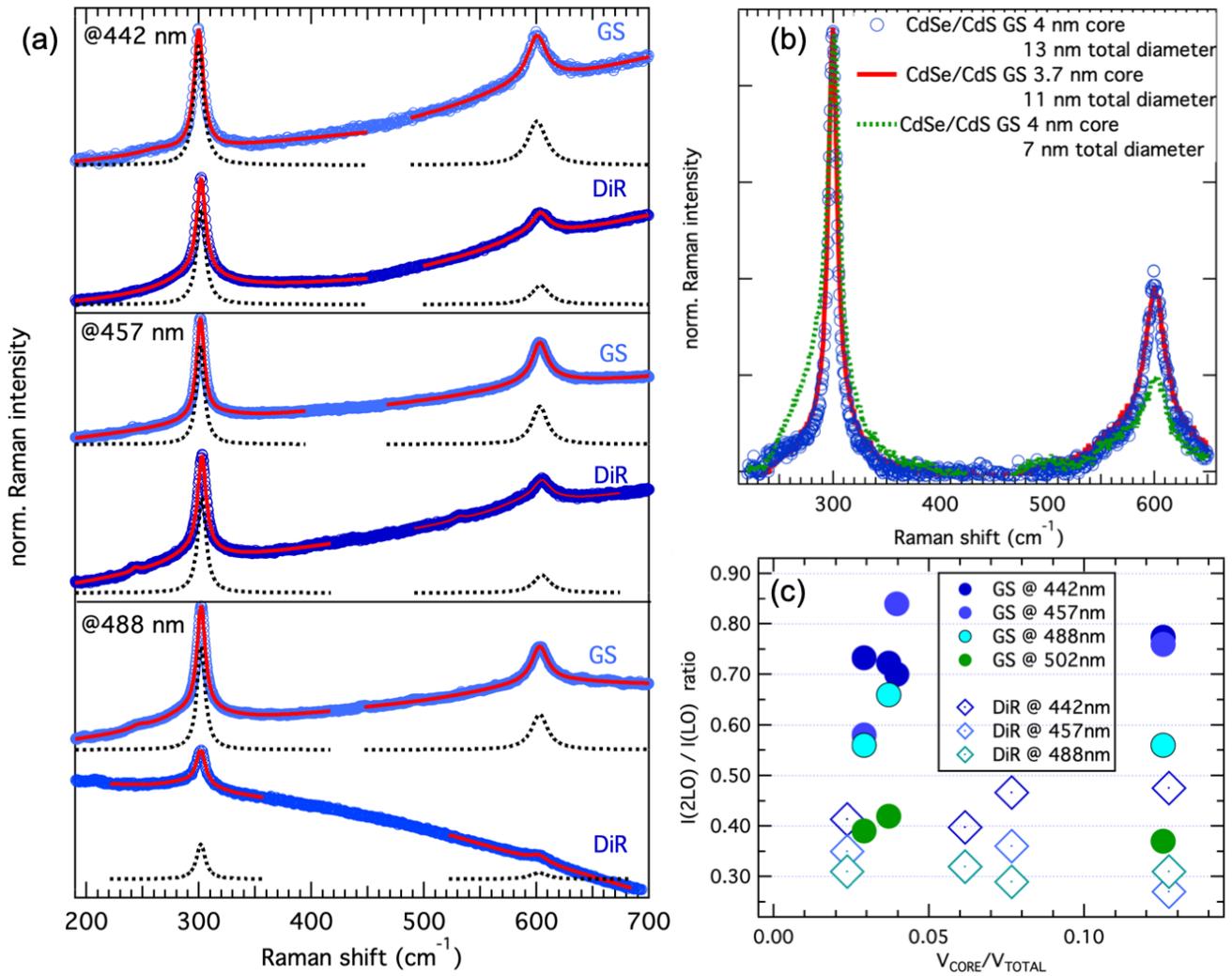

**Figure 2.** (a) Raman spectra recorded at 442, 457, and 488 nm excitation wavelength from GS and DiR nanocrystals that show pronounced LO and 2LO resonances. The Lorentz fits to the LO and 2LO peaks are reported by the black dotted lines, and the overall fitting is superposed as a red line with the data. Additional fitting peaks are not shown for clarity. The GS nanocrystals have 13 nm diameter and 4 nm core size, and the DiR 4.8 nm diameter, 30 nm length, and 4 nm core size. In all cases, the 2LO peak is significantly stronger for the GS nanocrystals with respect to the DiRs. (b) Resonant Raman spectra of GS CdSe/CdS nanocrystals with similar core diameter of ca. 4 nm, and different shell thickness, excited at 442 nm. For nanocrystal diameters of 11 nm and 13 nm the Raman spectra are very similar, however, for 7 nm diameter the 2 LO phonon replica is significantly weaker, and the LO peak is more broadened towards lower frequencies. (c) I(2LO)/I(LO) of the fitted Lorentz peaks versus the volume ratio of the core with respect to the overall nanocrystal ($V_{CORE}/V_{TOTAL}$) for the different excitation wavelengths.

The CdSe and CdS LO phonon peaks of the GS nanocrystals can be detected by nonresonant Raman spectroscopy as well, as shown in Figure 3a. However, under such nonresonant conditions, at 785 nm excitation wavelength, the peaks corresponding to multiphonon scattering are very weak, and the LO phonon replicas can hardly be identified. One possible reason is that such Raman scattering with excitation well below the band gap probes the excitonic levels at the heteronanocrystal band edge *via* virtual transitions, and that the extremely short timescales related to these processes hamper multiscattering events. Concerning the acoustic phonons, the RBMs and a set of overtones is evident in Figure 3b. Clearly, acoustic phonon modes can be distinguished that can be related to the RBMs of the spherical nanocrystals. The ground mode and first two overtone RBMs are labeled RBM1, RBM2, and RBM3 in Figure 3b, and can be well explained by linear elastic theory using Lamb's approach.[46] Therefore, contrary to the multiphonon scattering replicas of the LO phonons, the frequencies of the higher order RBMs correspond to higher vibrational harmonics of the nanocrystals, and therefore most likely do not originate from multiphonon scattering events. Furthermore, we observe a peak with a much smaller FWHM at 42 cm$^{-1}$, whose frequency does not shift with the nanocrystal diameter. This peak can be associated to the $TO_{\Gamma 6}$ mode of CdS [47]. The frequency of the RBMs scales with 1/D and can be well fitted with the equation $\omega_{RBM,n} = S_n \frac{2 c_L}{D}$ .[46] Here $S_n$ are coefficients, $D$ is the nanocrystal diameter and $c_L$ is the sound velocity that is used as fitting parameter. The fits to the data points of RBM1 and RBM2 in Figure 3b yield $3.45 \times 10^5 \, cm/s$ and $c_{RBM2} = 3.13 \times 10^5 \, cm/s$ as sound velocities, respectively. These values are considerably lower than the bulk sound velocity in CdS (4.3× 10$^5$ cm/s). Partially, this can be assigned to the softer CdSe core that leads to a lower overall sound velocity in the CdSe/CdS nanocrystals.[48] Furthermore, reduced sound velocities are commonly observed in colloidal nanocrystals and the reduction can be related to damping of the vibration due to the surface ligands. [49] Therefore, the different stiffness of the two materials should result in an average sound velocity that, together with the surface damping, determines the frequency of the oscillation. In this scenario, it is reasonable that the sound velocity obtained from the fitting of the RBM2 is lower than that related to RBM1, since the strain potential associated to the vibration of the RBM2 is localized closer to the surface of the nanocrystal.

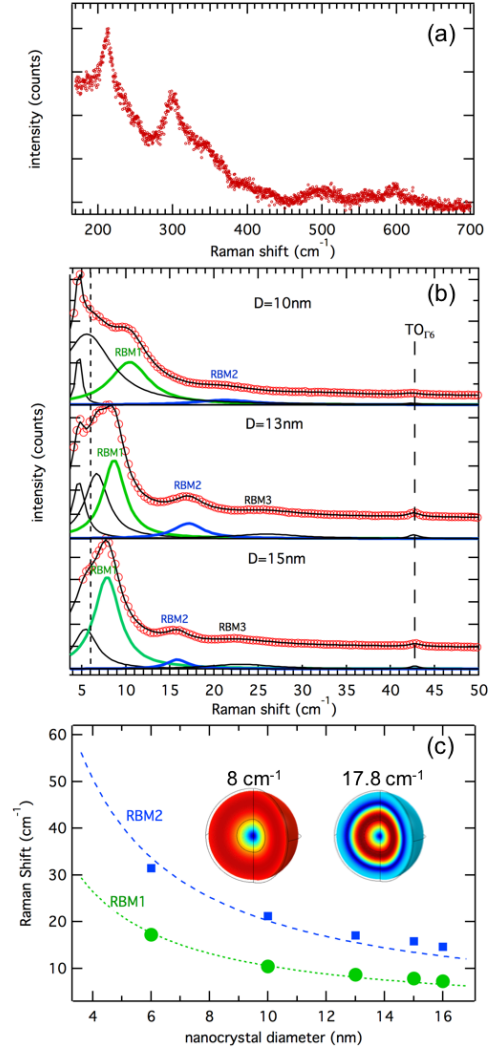

**Figure 3.** (a,b) Nonresonant Raman spectra of GS nanocrystals (10 nm diameter) with excitation wavelength at 785 nm. (a) The LO phonon bands of CdSe (around 210 cm-1) and CdS (around 300 cm$^{-1}$) are present with equal intensity, while their multiscattering replicas around 420 cm$^{-1}$ and 600 cm$^{-1}$ can only be faintly recognized. (b) Nonresonant Raman spectra of GS nanocrystals with different diameter and core size, recorded with 785 nm excitation wavelength. The radial breathing modes (RBMs) are highlighted by the green fits, and their first overtone by the blue fits. (b) Frequency of the RBM1 and RBM2 vs. nanocrystal diameter with the Lamb's theory fitting. The RBM frequency was fitted with $\omega_{RBM} = S_i \ c/D$, (with $S_1$=0.92; $S_2$=1.96).[46] The inset in (c) shows finite elements (COMSOL) calculations

of the strain potential for the first (RBM1) and second (RBM2) breathing mode oscillation for a core-shell CdSe/CdS nanocrystal with 4 nm core diameter and 14 nm overall diameter.

In conclusion, we have demonstrated by resonant Raman spectroscopy that the shape and architecture of CdSe/CdS core-shell heteronanocrystals has a strong impact on the exciton-phonon coupling. In particular, for excitation above the CdS band edge, the spherical GS nanocrystals showed a consistently stronger exciton-phonon coupling as compared to rod-shaped DiRs. Such strong coupling in GS nanocrystals should favor the rapid relaxation of high energy excitons to the band edge by efficient dissipation of their energy to optical phonons. Weaker coupling in spherical nanocrystals with a comparatively thin shell outlines the transition from "normal" to GS core-shell nanocrystals. Furthermore, low-frequency acoustic phonons are also prominent in GS nanocrystals, which can facilitate transitions within the fine structure of the exciton band edge state.[50] Therefore our study on the exciton phonon coupling in different core-shell architectures underpins the favorable properties of GS nanocrystals for light emission where with high quantum yield, amplified spontaneous emission and lasing are the key requisites.

ACKNOWLEDGMENT

We acknowledge support from the National Key Research and Development Program of China (Grant No. 2016YFA0301204), the National Natural Science Foundation of China (Grant Nos. 11474277, 11874350 and 11434010) and from K. C. Wong Education Foundation. And the research leading to these results has received funding from Horizon 2020 under the Marie Skłodowska-Curie Grant Agreement COMPASS No. 691185. B.M.G. acknowledges funding from the European Union's Horizon 2020 research and innovation program under grant agreement no.785219-GrapheneCore2.

TOC GRAPHIC

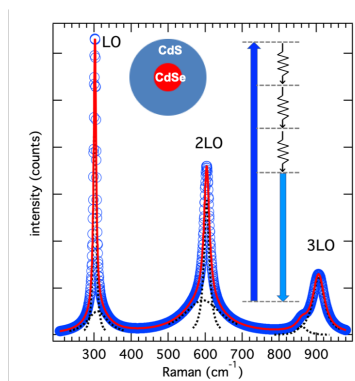